\documentclass[prd,twocolumn,showpacs,preprintnumbers,amsmath,amssymb,superscriptaddress,floatfix,nofootinbib]{revtex4-2}

\usepackage{graphicx}
\usepackage{bm}
\usepackage{amsmath}
\usepackage{amsfonts}
\usepackage{amssymb}
\usepackage{color}
\usepackage{subfigure}
\usepackage{multirow}
\usepackage{footnote}

\usepackage[colorlinks, citecolor=blue,anchorcolor=red,menucolor=red, linkcolor=red,filecolor=red,runcolor=red,urlcolor=blue,frenchlinks=red]{hyperref}
\begin{document}
\title{Strangeonium spectrum with the screening effects and interpretation of $h_1(1911)$ and $X(2300)$ observed by BESIII}

\author{Wei Hao}
\email{haowei@nankai.edu.cn}
\affiliation{School of Physics, Zhengzhou 	University, Zhengzhou 450001, China}

\author{M. Atif Sultan}%
\email{atifsultan.chep@pu.edu.pk}
\affiliation{School of Physics, Nankai University, Tianjin 300071, China}
\affiliation{Centre  For  High  Energy  Physics,  University  of  the  Punjab,  Lahore  (54590),  Pakistan}

\author{Li-Juan Liu}
\email{liulijuan@zzu.edu.cn}
\affiliation{School of Physics, Zhengzhou 	University, Zhengzhou 450001, China}

\author{En Wang}
\email{wangen@zzu.edu.cn}
\affiliation{School of Physics, Zhengzhou 	University, Zhengzhou 450001, China}

\begin{abstract}
Motivated by two news states $h_1(1911)$ and $X(2300)$ observed by BESIII, we have investigated the mass spectrum and the strong decay properties of the strangeonium mesons within the modified Godfrey-Isgur model by considering the screening effects. We have determined the free parameters using the masses and widths of the well established $s\bar{s}$ states $\phi(1020)$, $\phi(1680)$, $h_1(1415)$, $f_2^\prime(1525)$, and $\phi_3(1850)$. According to our results, $h_1(1911)$ and $X(2300)$ could be well explained as states $h_1(2^1P_1)$ and $h_1(3^1P_1)$ $s\bar{s}$ states, respectively.  Meanwhile, the possible assignments of $X(2000)$, $\eta_2(1870)$, and $\phi(2170)$ as $3^3S_1$, $1^1D_2$, and $2^3D_1$ are also discussed. Furthermore, the masses and widths of the $2S$, $3S$, $1P$, $2P$, $3P$, $1D$, and $2D$ $s\bar{s}$ states are also given and compared with various theoretical predictions, which is helpful for the observations and confirmations of these states in future.
\end{abstract}

\maketitle

\section{Introduction}
The strangeonium states lying between the light $q\bar{q}$ states and heavy $Q\bar{Q}$ states are related to the long-range
interactions, and studying the $s\bar{s}$-meson system is crucial to deepening our understanding of the non-perturbative properties of Quantum
Chromodynamics (QCD)~\cite{Liu:2015zqa,Ketzer:2019wmd,Brambilla:2019esw,vanBeveren:2020eis,Li:2020xzs,Godfrey:1985xj,Burakovsky:1997dd,Pang:2019ttv,Gao:2019jme,Cheng:2011pb,Burakovsky:1997ci,Li:2005eq}. However, for a long time, only several strangeonium states have been well established, i.e. $\phi(1020)$, $\phi(1680)$, $h_1(1415)$, $f_2^\prime(1525)$, and $\phi_3(1850)$~\cite{Li:2020xzs}. Meanwhile, the mixing between the singlet and the octet caused by the SU(3) flavor symmetry breaking also complicates our understanding of the strangeonium states. Systematically studying of the $s\bar{s}$ states can help distinguish between traditional $q\bar{q}$ states and non-$q\bar{q}$ states.
\par
As a powerful platform to study the $s\bar{s}$ state, the BESIII Collaboration has performed a partial wave analysis of the process $J/\psi\to \phi \pi^0\eta$ in 2023, and  observed two new states $h_1(1900)$ and $X(2000)$ with statistical significance of $24.0\sigma$ and $16.9\sigma$ in the $\phi \eta$ invariant mass distribution~\cite{BESIII:2023zwx}. The quantum numbers, mass, and width of $h_1(1900)$ are $J^{PC}=1^{+-}$, $M=1911\pm6\pm14$~MeV, and $\Gamma=149\pm12\pm23$~MeV, and those of $X(2000)$ are $J^{PC}=1^{--}$, $M=1996\pm11\pm30$~MeV, and $\Gamma=148\pm16\pm66$~MeV. 
\par
It should be pointed that another state with mass around 2.0 GeV was also observed  in the $\phi\eta^\prime$ mass spectrum of the process $J/\psi\to \phi\eta\eta^\prime$ by BESIII in 2019~\cite{BESIII:2018zbm}. Its mass and width are given by $M=2002.1\pm27.5\pm15.0$~MeV and $\Gamma=129\pm17\pm7$~MeV when assuming the spin-parity quantum numbers $J^P=1^-$ for this state. However, when regarding the state as $J^P=1^+$, the mass and width will be $M=2062.8\pm 13.1\pm4.2$~MeV and $\Gamma=177\pm36\pm20$~MeV.  Since the mass and width of this state observed in Ref.~\cite{BESIII:2018zbm} are in consistent with the ones of $X(2000)$ observed in Ref.~\cite{BESIII:2023zwx} within the experimental uncertainties, two states could be the same state. 
\par
Furthermore, recently BESIII has also reported another new state $X(2300)$ by a partial wave analysis in the process $\psi(3686)\to \phi\eta\eta^\prime$ with covariant tensor approach, and determined the mass $M=2316\pm9\pm30$~MeV and width $89\pm15\pm26$~MeV~\cite{BESIII:2024nhv}. These newly observed states $h_1(1900)$, $X(2000)$, and  $X(2300)$ further enrich the $s\bar{s}$ spectrum. 
\par
The spectrum of the $s\bar{s}$ states have been studied  in many works. For instance, in Ref.~\cite{Pang:2019ttv}, by using modified Godfrey and Isgur model  (MGI model), the $\phi$ family has been well studied. In this work the new observed state X(2000) in 2019 is explained as the $\phi(3S)$, and the state $\phi(2170)$ [also named $Y(2175)$] is assigned as the $\phi(2D)$ state. Furthermore, some states such as $f_2(2010)$, $f_2(2150)$, $X(2060)$, and $X(2500)$ can also been explained as strangeonium states based on a nonrelativistic linear potential quark model~\cite{Li:2020xzs}. The spectrum of the $s\bar{s}$ states, especially the excited states have not yet well established, and the recently observed states could be shed light on this  issue.
\par
As we know, many potential models have been developed to study hadron spectrum~\cite{Brown:1979ya,Eichten:1980mw,Gupta:1981pd,Pantaleone:1985uf,Godfrey:1985xj,Lakhina:2006fy}. In these models, the potential between quarks usually includes a Coulomb term at short distance and a linear confining term at large distance. Due to the fact that these models explain hadron states based on the fundamental assumptions of $q\bar{q}$ and $qqq$, without coupling with other components, they are commonly referred to as quenched quark models. However, as more and more hadron states are discovered in experiments, many of them can no longer be well explained by quenched quark models. Except for developing some exotic state models such as multiquark state models, molecular state models, and $et$ $al$, the unquenched model is also a good candidate in this direction. The unquenched effects could come from the contributions of the fluctuation caused by sea quarks and the vacuum polarization effects. The previous one causes the coupled channel model~\cite{Lu:2016mbb,Ferretti:2012zz,Ferretti:2013faa,Ortega:2016mms,Hao:2022vwt,Yang:2023tvc,Hao:2024nqb,Hao:2024ptu}, and the latter causes the screened potential model~\cite{Laermann:1986pu,Born:1989iv,Armoni:2008jy,Li:2009ad,Hao:2019fjg,Feng:2022esz,Hao:2022ibj,Song:2015fha,Song:2015nia,Wang:2019mhs,Wang:2018rjg}. These effects can cause mass shifts with respect to the predictions of the quenched quark models. In fact, both the coupled channel model and the screened potential model have similar features under certain conditions~\cite{Li:2009ad}. The MGI model is developed by combining the GI model and screening effects~\cite{Song:2015fha,Song:2015nia,Wang:2019mhs,Wang:2018rjg}, where the linear potential in GI model is replaced with a screened potential, which is important for high excited states.
\par
Thus, inspired by new observed states, we will systematically study the spectrum and decay properties of the strangeonium by using the  MGI model and the $^3P_0$ model. 
We will investigate the possible assignments of the $s\bar{s}$ states, especially $h_1(1911)$ and $X(2300)$. Furthermore, we will calculate the masses and strong decay properties of other $S$, $P$, and $D$-wave mesons, and  discuss the possible assignments of the states $X(2000)$, $\eta(1870)$, and $\phi(2170)$.

 Recently, we noted that Ref.~\cite{Wang:2024lba} has studied the mass spectrum and OZI-allowed two-body decays of axial-vector mesons, including $X(2300)$, $b_1$, $h_1$, and $h_1^\prime$. They computed the mass spectrum and wave functions by diagonalizing the Hamiltonian matrix using a harmonic oscillator basis. Their conclusion that $h_1(1911)$ can be assigned as $h_1(2^1P_1)$ aligns with our results. However, while we interpret $X(2300)$ as $h_1(3^1P_1)$, its mass and width remain unexplained in Ref.~\cite{Wang:2024lba}.

\par
This article is organized as follows: In  Sec.~\ref{sec:model}, we give a brief introduction of the modified GI model and the $^3P_0$ model. In Sec.~\ref{sec:results}, the numerical results are presented. The summary is given in Sec.~\ref{sec:summary}.

\section{Formalism}
\label{sec:model}

\subsection{Modified Godfrey-Isgur model}
\label{sec:MGI}

In this section, we briefly introduce the modified GI (MGI) model. Firstly, for the GI model, the Hamiltonian of a meson system with $q\bar{q}$ is defined as~\cite{Godfrey:1985xj}
\begin{eqnarray}
\tilde{H}&=&\left(p^2+m_q^2\right)^{1/2}+\left(p^2+m_{\bar{q}}^2\right)^{1/2}+V_\mathrm{eff}(\mathbf{p,r}),
\end{eqnarray}
with
\begin{eqnarray}
V_\mathrm{eff}(\mathbf{p,r})=\tilde{H}_{q\bar{q}}^{\text{conf}}+\tilde{H}_{q\bar{q}}^{\text{so}}+\tilde{H}_{q\bar{q}}^{\text{hyp}}.
\end{eqnarray}
where $\tilde{H}^{conf}_{q\bar{q}}$ is spin-independent potential, $\tilde{H}^{hyp}_{q\bar{q}}$ is color-hyperfine interaction, $\tilde{H}^{so}_{q\bar{q}}$ is spin-orbit interaction. The explicit expression of $\tilde{H}^{conf}_{q\bar{q}}$, $\tilde{H}^{hyp}_{q\bar{q}}$, and $\tilde{H}^{so}_{q\bar{q}}$ are given in Refs.~\cite{Song:2015nia,Li:2021qgz,Feng:2022hwq}.

The spin-independent potential $\tilde{H}^{conf}_{q\bar{q}}$ contains a constant term, a linear confining potential, and a one-gluon exchange potential
\begin{equation}
\tilde{H}^{conf}_{q\bar{q}}=\widetilde{G}(r)+\widetilde{S}(r),
\end{equation}
\begin{equation}
\widetilde{G}(r)=-\sum_k \frac{4 \alpha_k}{3 r}\left[\frac{2}{\sqrt{\pi}} \int_0^{\tau_{k i j} r} e^{-x^2} d x\right],
\end{equation}
\begin{eqnarray}
\widetilde{S}(r)&=&b r {\left[\frac{e^{-\sigma_{i j}{ }^2 r^2}}{\sqrt{\pi} \sigma_{i j} r}\right.} \nonumber \\
&&\left.+\left[1+\frac{1}{2 \sigma_{i j}{ }^2 r^2}\right] \frac{2}{\sqrt{\pi}} \int_0^{\sigma_{i j}{ }^r} e^{-x^2} d x\right]+c,
\end{eqnarray}
where
\begin{equation}
\frac{1}{\tau_{k i j}{ }^2}=\frac{1}{\gamma_k{ }^2}+\frac{1}{\sigma_{i j}{ }^2} .
\end{equation}

By considering the strengths of the various interactions depending on the center-of-mass (C.M.)  momentum of the interactions quarks, the potentials should be revised by the momentum-dependent factors. Then the Coulomb term, the contact, tensor, vector spin-orbit, and scalar spin-orbit potentials should be modified according to
\begin{eqnarray}
\tilde{G}(r) &\to& \left(1+\frac{p^2}{E_1 E_2}\right)^{1/2} \tilde{G}(r) \left( 1+\frac{p^2}{ E_1 E_2}\right)^{1/2},\nonumber\\
\frac{\tilde{V}_i(r)}{m_1 m_2} &\to& \left( \frac{m_1 m_2}{E_1 E_2}\right)^{1/2+\epsilon_i}  \frac{\tilde{V}_i(r)}{m_1 m_2} \left( \frac{m_1 m_2}{E_1 E_2}\right)^{1/2+\epsilon_i}.\label{eqGV}
\end{eqnarray}
where $\epsilon_i$ is the parameter of each term, $i$ corresponds to the  contact($c$), tensor($t$), vector spin-orbit[so($v$)], and scalar spin-orbit[so($v$)] terms\cite{Godfrey:1985xj}. 

Now, the effective potential of the Hamiltonian is expressed as

\begin{widetext}\begin{eqnarray}
G_{\mathrm{eff}}(r) &=& \left(1+\frac{p^2}{E_1 E_2}\right)^{1 / 2} \widetilde{\boldsymbol{G}}(r)\left(1+\frac{p^2}{E_1 E_2}\right)^{1 / 2} \nonumber \\
&& +\left[\frac{\mathbf{S}_1 \cdot \mathbf{L}}{2 m_1{ }^2} \frac{1}{r} \frac{\partial \widetilde{G}_{11}^{\mathrm{so}(v)}}{\partial r}+\frac{\mathbf{S}_2 \cdot \mathbf{L}}{2 m_2{ }^2} \frac{1}{r} \frac{\partial \widetilde{G}_{22}^{\mathrm{so}(v)}}{\partial r}+\frac{\left(\mathbf{S}_1+\mathbf{S}_2\right) \cdot \mathbf{L}}{m_1 m_2} \frac{1}{r} \frac{\partial \widetilde{G}_{12}^{\mathrm{so}(v)}}{\partial r}\right] \nonumber \\
&& +\frac{2 \mathbf{S}_1 \cdot \mathbf{S}_2}{3 m_1 m_2} \nabla^2 \widetilde{\boldsymbol{G}}_{12}^c-\left(\frac{\mathbf{S}_1 \cdot \hat{r}_2 \cdot \hat{r}-\frac{1}{3} \mathbf{S}_1 \cdot \mathbf{S}_2}{m_1 m_2}\right)\left(\frac{\partial^2}{\partial r^2}-\frac{1}{r} \frac{\partial}{\partial r}\right) \widetilde{\boldsymbol{G}}_{12}^t,
\end{eqnarray}
\end{widetext}
with
\begin{eqnarray}
S_{\text {eff }}(r)=\widetilde{S}(r)-\frac{\mathbf{S}_1 \cdot \mathbf{L}}{2 m_1^2} \frac{1}{r} \frac{\partial \widetilde{S}_{11}^{\mathrm{so}(s)}}{\partial r}-\frac{\mathbf{S}_2 \cdot \mathbf{L}}{2 m_2^2} \frac{1}{r} \frac{\partial \mathbf{S}_{22}^{\mathrm{so}(s)}}{\partial r}.
\end{eqnarray}
Where $\mathbf{L}$ and $\mathbf{S}$ are the orbit and spin quantum numbers. The subscripts 1 and 2 in the above expression correspond to the quark and antiquark of a meson.
\par
The modifition of the GI model is achieved by introducing the screening effects into the initial GI model, which can be taken into account by making a replacement as,
\begin{equation}
br\to V^{scr}(r)= \frac{b(1-e^{\mu r})}{\mu}.
\end{equation}
With this modification, the final expression for $\tilde{V}^{scr}(r)$ is shown as~\cite{Song:2015nia}
\begin{widetext}
\begin{eqnarray}
\tilde{V}^{\mathrm{scr}}(r)= & \frac{b}{\mu r}\left[r+e^{\frac{\mu^2}{4 \sigma^2}+\mu r} \frac{\mu+2 r \sigma^2}{2 \sigma^2}\left(\frac{1}{\sqrt{\pi}} \int_0^{\frac{\mu+2 r \sigma^2}{2 \sigma}} e^{-x^2} d x-\frac{1}{2}\right)\right. \nonumber \\
& \left.-e^{\frac{\mu^2}{4 \sigma^2}-\mu r} \frac{\mu-2 r \sigma^2}{2 \sigma^2}\left(\frac{1}{\sqrt{\pi}} \int_0^{\frac{\mu-2 r \sigma^2}{2 \sigma}} e^{-x^2} d x-\frac{1}{2}\right)\right].
\end{eqnarray}
\end{widetext}

\subsection{The $^3P_0$ model}
The two-body Okubo-Zweig-Iizuka (OZI) allowed strong decay behaviors are discussed by the $^3P_0$ model which is first proposed by L. Micu in 1969~\cite{Micu:1968mk} and then developed by A. Le Yaouanc et al. in the 1970s~\cite{LeYaouanc:1972vsx,LeYaouanc:1973ldf}. At present, it has been developed as an important tool for studying the strong decay properties of hadrons \cite{Hao:2022ibj,Li:2009rka,Pan:2016bac,Lu:2016bbk,Wang:2017pxm,Xue:2018jvi}.
In the $^3P_0$ model, the $S$ matrix of the decay $A\rightarrow BC$ is defined as
\begin{eqnarray}
  S = I-2\pi i\delta(E_f-E_i)T.
\end{eqnarray}
The transition operator $T$ in above expression is shown as \cite{Ferretti:2013faa,Ferretti:2012zz,Ferretti:2013vua}
\begin{eqnarray}
  T &=&-3\gamma\sum_m\langle 1m1-m|00\rangle\int
    d^3\boldsymbol{p}_3d^3\boldsymbol{p}_4\delta^3(\boldsymbol{p}_3+\boldsymbol{p}_4)\nonumber\\
&& \times   {\cal{Y}}^m_1\left(\frac{\boldsymbol{p}_3-\boldsymbol{p}_4}{2}\right
    )\chi^{34}_{1,-m}\phi^{34}_0\omega^{34}_0b^\dagger_3(\boldsymbol{p}_3)d^\dagger_4(\boldsymbol{p}_4),
\end{eqnarray}
where the indicator 3 and 4 in above formula correspond to the quark pairs generated by vacuum and its strength is represent by the dimensionless parameter $\gamma$.  The $\boldsymbol{p}_3$ and $\boldsymbol{p}_4$ are the momenta of the quark $q_3$ and $\bar{q}_4$, respectively. $\chi^{34}_{1,-m}$, $\phi^{34}_0$, and $\omega^{34}_0$ are the spin, flavor and color wave functions of the $q_3\bar{q}_4$, respectively. The solid harmonic polynomial ${\cal{Y}}^m_1(\boldsymbol{p})\equiv|p|^1Y^m_1(\theta_p, \phi_p)$ represents the momentum-space distribution of $q_3\bar{q}_4$. It should be pointed out that the meson space wave functions of the initial state $A$, and finial states $BC$ are obtained by solving the MGI model in our calculations.
\par
The partial wave amplitude ${\cal{M}}^{LS}(\boldsymbol{P})$ of the decay  $A\rightarrow BC$ can be given by~\cite{Jacob:1959at}
\begin{eqnarray}
{\cal{M}}^{LS}(\boldsymbol{P})&=&
\sum_{\renewcommand{\arraystretch}{.5}\begin{array}[t]{l}
\scriptstyle M_{J_B},M_{J_C},\\\scriptstyle M_S,M_L
\end{array}}\renewcommand{\arraystretch}{1}\!\!
\langle LM_LSM_S|J_AM_{J_A}\rangle \nonumber\\
&&\langle
J_BM_{J_B}J_CM_{J_C}|SM_S\rangle\nonumber\\
&&\times\int
d\Omega\,Y^\ast_{LM_L}{\cal{M}}^{M_{J_A}M_{J_B}M_{J_C}}
(\boldsymbol{P}), \label{pwave}
\end{eqnarray}
where ${\cal{M}}^{M_{J_A}M_{J_B}M_{J_C}}
(\boldsymbol{P})$ is the helicity amplitude and defined as,
\begin{eqnarray}
\langle
BC|T|A\rangle=\delta^3(\boldsymbol{P}_A-\boldsymbol{P}_B-\boldsymbol{P}_C){\cal{M}}^{M_{J_A}M_{J_B}M_{J_C}}(\boldsymbol{P}).\nonumber \\
\end{eqnarray}
The $|A\rangle$, $|B\rangle$, and $|C\rangle$ denote the mock meson states defined in Ref.~\cite{Hayne:1981zy}.
\par
With above amplitude, the decay width $\Gamma(A\rightarrow BC)$  can be obtained by
\begin{eqnarray}
\Gamma(A\rightarrow BC)= \frac{\pi
|\boldsymbol{P}|}{4M^2_A}\sum_{LS}|{\cal{M}}^{LS}(\boldsymbol{P})|^2, \label{width1}
\end{eqnarray}
where $|\boldsymbol{P}|=\frac{\sqrt{[M^2_A-(M_B+M_C)^2][M^2_A-(M_B-M_C)^2]}}{2M_A}$,
and $M_A$, $M_B$, and $M_C$ are the masses of the meson $A$, $B$, and $C$, respectively. 

\section{Results and discussions}
\label{sec:results}

\begin{table}[htpb] 
\caption{Parameters of the modified Godfrey-Isgur model.} 
\label{tab:gipar}
\begin{center}
\begin{tabular}{ccc} 
\hline 
\hline 
Parameter             &  Value \\ 
\hline 
$m_u$                 &0.22        \\
$m_d$                 &0.22        \\
$m_s$                 &0.419        \\
$b$                   &0.19318  GeV$^2$    \\  
$c$                   &-0.249  GeV       \\  
$\sigma_0$            &1.8401  GeV        \\
$s$                   &1.554             \\
$\epsilon_c$          &-0.168             \\
$\epsilon_t$          &0.025            \\
$\epsilon_{\mathrm{so(v)}}$   &-0.035       \\
$\epsilon_{\mathrm{so(s)}}$   &0.05       \\  
$\mu$                 &0.05      \\
\hline 
\hline
\end{tabular}
\end{center}
\end{table}

\begin{table*}[htpb]
\begin{center}
\caption{\label{tab:mass} The mass spectrum (in MeV) of the strangeonium. }
\footnotesize
\begin{tabular}{cccccccccccc}
\hline\hline
  $n^{2S+1}L_J$  & states               & RPP~\cite{ParticleDataGroup:2024cfk}    &This work &QL\cite{Li:2020xzs}  &GI\cite{Godfrey:1985xj}   &XWZZ\cite{Xiao:2019qhl}  &EFG\cite{Ebert:2009ub}  &SIKY\cite{Ishida:1986vn}  &Pang\cite{Pang:2019ttv}  &VFV\cite{Vijande:2004he} \\\hline
  $1^1S_0$       &                      &                        &657       &797  &960  &657   &743  &690   &      &956\\
  $1^3S_1$     & $\phi(1020)$           &$1019.461\pm0.016$      &1020      &1017 &1020 &1009  &1038 &1020  &1030  &1020\\
  $2^1S_0$     &                        &                        &1542      &1619 &1630 &1578  &1536 &1440  &      &1795  \\
  $2^3S_1$     & $\phi(1680)$           &$1680\pm20$             &1646      &1699 &1690 &1688  &1698 &1740  &1687  &1726   \\
  $3^1S_0$     &                        &                        &2022      &2144 &     &2125  &2085 &1970  &      &    \\
  $3^3S_1$     & $X(2000)$              &$1996\pm11\pm30$        &2092      &2198 &     &2204  &2119 &2250  &2147  &   \\
  $1^3P_0$     & $f_0(1370)$            &$1200\to1500$           &1340      &1373 &1360 &1355  &1420 &1180  &      &   \\
  $1^1P_1$     & $h_1(1415)$            &$1409^{+9}_{-8}$        &1460      &1462 &1470 &1473  &1485 &1460  &      &1511    \\
  $1^3P_1$     & $f_1(1420)$            &$1426.3\pm0.9$          &1464      &1492 &1480 &1480  &1464 &1430  &      &1508    \\
  $1^3P_2$     & $f_2^\prime(1525)$     &$1517.4\pm2.5$          &1523      &1513 &1530 &1539  &1529 &1480  &      &1556    \\
  $2^3P_0$     &                        &                        &1907      &1971 &1990 &1986  &1969 &1800  &      &  \\
  $2^1P_1$     & $X(1911)$              &$1911\pm6\pm14$         &1934      &1991 &2010 &2008  &2024 &2040  &      &1973 \\
  $2^3P_1$     &                        &                        &1948      &2027 &2030 &2027  &2016 &2020  &      &   \\
  $2^3P_2$     &                        &                        &1969      &2466 &     &2480  &2412 &2540  &      &   \\
  $3^3P_0$     &                        &                        &2290      &2434 &     &2444  &2364 &2280  &      & \\
  $3^1P_1$     & $X(2300)$            &$2316\pm9\pm31$         &2301      &2435 &     &2449  &2398 &2490  &      &  \\
  $3^3P_1$     &                        &                        &2314      &2470 &     &2468  &2403 &2480  &      &  \\
  $3^3P_2$     &                        &                        &2329      &2466 &     &2480  &2412 &2540  &      &    \\
  $1^3D_1$     &                        &                        &1832      &1809 &1880 &1883  &1845 &1750  &1869  & \\
  $1^1D_2$     & $\eta_2(1870)$         &$1842\pm8$              &1848      &1825 &1890 &1893  &1909 &1830  &      &1853 \\
  $1^3D_2$     &                        &                        &1857      &1840 &1910 &1904  &1908 &1810  &      &   \\
  $1^3D_3$     & $\phi_3(1850)$         &$1854\pm7$              &1853      &1822 &1900 &1897  &1950 &1830  &      &1875  \\
  $2^3D_1$     & $\phi(2170)$           &$2164\pm6$              &2220      &2272 &     &2342  &2258 &2260  &2276  &  \\
  $2^1D_2$     &                        &                        &2220      &2282 &     &2336  &2321 &2340  &      &   \\
  $2^3D_2$     &                        &                        &2229      &2297 &     &2348  &2323 &2330  &      &   \\
  $2^3D_3$     &                        &                        &2223      &2285 &     &2337  &2338 &2360  &      &  \\
  \hline\hline

\end{tabular}
\end{center}
\end{table*}

\begin{table*}[htpb]
\begin{center}
\caption{\label{tab:decay1} The decay widths (in MeV) of strangeonium. The data in \{\} means the width is calculated by using experimental initial state mass. }
\footnotesize
\begin{tabular}{cccccccccccccc}
\hline\hline
$n^{2S+1}L_J$      &$1^3S_1$ &$2^1S_0$ &$2^3S_1$  &$1^3P_0$   &$1^1P_1$       &$1^3P_1$        &$1^3P_2$    \\\hline
$K\bar{K}$         &0.8\{0.8\} &$-$      &15.9\{15.2\}  &238.4  &$-$            &$-$             &20.9\{20.3\} \\
$K^*\bar{K}$       &$-$      &147.3    &114.4\{127.7\}&$-$    &138.6\{81.6\}  &278.0\{210.5\}  &5.9\{5.3\}  \\
$\eta\eta$         &$-$      &$-$      &$-$           &17.1   &$-$            &$-$             &0.7\{0.7\} \\
$\eta\eta^\prime$  &$-$      &$-$      &$-$           &$-$    &$-$            &$-$             &0.0007\{0.0003\} \\
$\eta \phi$        &$-$      &$-$      &2.4\{4.0\}    &$-$    &$-$            &$-$             &$-$    \\
Total              &0.8\{0.8\}   &147.3    &132.7\{146.9\}&255.5  &138.6\{81.6\}  &278.0\{210.5\}  &27.5\{26.3\}   \\
Exp.               &$4.249\pm0.013$ &  &$150\pm50$    &$200\to500$ &$78\pm11$ &$54.5\pm2.6$    &$86\pm5$ \\

  \hline\hline
\end{tabular}
\end{center}
\end{table*}

In this work, only the pure $s\bar{s}$ states are investigated. For strangeonium, we use $\phi(1020)$, $\phi(1680)$, $h_1(1415)$, $f_2^\prime(1525)$, and $\phi_3(1850)$, which can be regarded as good candidates of $s\bar{s}$ states~\cite{Li:2020xzs}, to fit the model parameters.\footnote{
    It is worth noting that the $h_1(1595)$  shares the same quantum numbers with the $h_1(1415)$. While Ref.~\cite{Wang:2024lba} regards the $h_1(1595)$ as a candidate for the $2^1P_1$ $q\bar{q}$ ($q=u,d$), the partner of the $h_1(1900)$. We have re-evaluated this assignment by treating the $h_1(1595)$ as the $1^1P_1$ $s\bar{s}$ state. Our analysis shows that the predicted mass (1460 MeV) of the $1^1P_1$ $s\bar{s}$ state aligns more closely with the $h_1(1415)$ than with the $h_1(1595)$. Therefore, in this work, we adopt the $h_1(1415)$ as the $1^1P_1$ $s\bar{s}$ state, consistent with Refs.~\cite{Vijande:2004he,Li:2020xzs}.
    }
 However, these states are not enough to fit all model parameters. So we only fit a few parameters which have significant impact on the model's predictions, such as $b$, $c$, $\sigma$, $s$, and $\mu$.  All the parameters involved in the MGI model are listed in Table~\ref{tab:gipar}. With these parameters, the spectrum and the strong decay properties of strangeonium are tabulated in Table~\ref{tab:mass} and Table~\ref{tab:decay1}-\ref{tab:decay3}, respectively. There are discrepancies between our parameters and those used in Ref.~\cite{Wang:2024lba}. In Ref.~\cite{Wang:2024lba}, the parameters are fitted to the experimental data of the $1^{+-}$ states, while in this work, the parameters of the MGI model are fitted using established strangeonium states (e.g., $\phi(1020)$, $\phi(1680)$, $h_1(1415)$, $f_2^\prime(1525)$, and $\phi_3(1850)$) as benchmarks. Additionally, the exact uncertainties of the predictions are difficult to determine, though they are typically estimated to be around $20\sim 30$\%~\cite{Godfrey:2016nwn,Feng:2022esz}.

\subsection{$S$-wave $s\bar{s}$ states}
The lowest $S$-wave vector $s\bar{s}$ state is $\phi(1020)$, which is observed in a bubble chamber experiment at Brookhaven in 1962~\cite{Bertanza:1962zz}, which has been confirmed by large number of experiments. Its mass and width are given by $M=1019.461\pm0.016$~MeV and $\Gamma=4.249\pm0.013$~MeV~\cite{ParticleDataGroup:2024cfk}. The state can be well explained as $1^3S_1$~\cite{Li:2020xzs}. Our prediction of its mass is 1020~MeV which supports the assignment of $\phi(1020)$ as $1^3S_1$. 

\par
For the $2S$-wave state $s\bar{s}$, currently there is no experimental data available for the state $2^1S_0$, and our predicted mass is 1542~MeV, as shown in Table~\ref{tab:mass}, which is close to the predicted mass 1578~MeV based on the GI model~\cite{Xiao:2019qhl} and 1536 MeV based on Regge trajectories~\cite{Ebert:2009ub}. Compared with different model results, the mass range of the state $2^1S_0$ $s\bar{s}$ should be $1440\sim1630$ MeV, and its main decay mode is $K^*\bar{K}$ with width 147~MeV, as shown in Table~\ref{tab:decay1}. For the $2S$-wave vector meson, $\phi(1680)$ is a good candidate of $2^3S_1$. The state was first discovered in $e^+e^-\to K_SK^\pm\pi^\mp$ with an observed mass $1680\pm20$~MeV and a width $150\pm50$~MeV~\cite{Mane:1982si,ParticleDataGroup:2024cfk}. Our predicted mass and width can well match the experimental values, especially the mainly decay modes $KK$, $K^*\bar{K}$ and $\eta\phi$. 
\par 
By now, there is no observation of the $3^1S_0$ $s\bar{s}$ state in experiments. As shown in Table~\ref{tab:mass}, the predicted range mass is $1970\sim 2144$~MeV. Based on the $^3P_0$ model with wave functions obtained using the MGI model, we predict the width 158~MeV for this state, and the dominant decay modes are $K^*\bar{K}$, $K^*\bar{K}^*$, $K^*_0(1430)\bar{K}$, $K^*_2(1430)\bar{K}$, $\eta f_0(1370)$, and $\eta f_1(1420)$. For the $3^3S_1$ state, the predicted mass range is $2092\sim 2250$ MeV from various models as shown in Table~\ref{tab:mass}, and our predicted width is $118$ MeV. By comparing the mass and decay width, we support to assign $X(2000)$ [$M=1996\pm 11\pm30$~MeV and $\Gamma=148\pm 16\pm 66$~MeV] as $3^3S_1$ vector meson.

\subsection{$1P$-wave $s\bar{s}$ states}
The four $1P$-wave states are the states with the most abundant experimental information. For the states $1^3P_0$, $1^1P_1$, $1^3P_1$, and $1^3P_2$, our predicted  masses are 1340~MeV, 1460~MeV, 1464~MeV, and 1523~MeV, respectively, as shown in Table~\ref{tab:mass}.
\par
For the $f_0(1370)$, the amplitude analysis of the $K_SK_S$ system produced in the radiative decays of $J/\psi$ gives its mass  $M=1350\pm9^{+12}_{-2}$~MeV and width $\Gamma=231\pm21^{+28}_{-48}$~MeV~\cite{BESIII:2018ubj}. Recently, the coupled channel analysis of $\bar{p}p\to \pi^0\pi^0\eta, \pi^0\eta\eta$ and $K^+K^0\pi^0$ at 900~MeV/c and of $\pi\pi$-scattering data gives its $T$-matrix pole $(1280.6\pm1.6\pm47.4)-i(205.2\pm1.7\pm20.7)$~MeV~\cite{CrystalBarrel:2019zqh}. The study of the isoscalar scalar mesons and the scalar glueball from radiative decays of $J/\psi$ gives its $T$-matrix pole $(1370\pm40)-i(195\pm20)$~MeV~\cite{Sarantsev:2021ein}. These informations show that the $f_0(1370)$ mass has large uncertainty but its average mass value is about $1200\sim 1500$ MeV, which is consistent with our prediction within error bars. Besides, we have also calculated its strong decay properties in Table~\ref{tab:decay1}, and the predicted total width is $\Gamma =307$~MeV and the  dominant decay modes are $K\bar{K}$ and $\eta\eta$, which are in agreement with the experimental measurements. The ratio of the partial widths of two channels is $\Gamma_{\eta\eta}/\Gamma_{K\bar{K}}\approx 0.29$, which is close to the result 0.20 of Ref.~\cite{Li:2020xzs}.
\par
In 2018, the BESIII Collaboration has observed $h_1(1415)$ with mass $1423\pm2.1\pm7.3$~MeV and width $90.3\pm9.8\pm17.5$~MeV in the process $J/\psi\to\eta^\prime K\bar{K}\pi$~\cite{BESIII:2018ede}.
Later, BESIII has reported the $h_1(1415)$ in the partial wave analysis of $J/\psi\to \gamma\eta^\prime\eta^\prime$, and determined the mass and width to be $1384\pm6^{+9}_{-0}$~MeV and $66\pm10^{+12}_{-10}$~MeV~\cite{BESIII:2022zel}, respectively.  Theoretically,   the state $h_1(1415)$ can be explained as a dynamically generated resonance of the pseudoscalar-vector interaction~\cite{Jiang:2019ijx,Roca:2005nm}, a mixed state of a singlet and an octet with mixing angle $29.5^\circ$~\cite{Gao:2019jme}, or a pure $s\bar{s}$ state in the constituent quark model~\cite{Vijande:2004he,Li:2020xzs}. Our predicted mass 1460~MeV is larger than the experimental data but consistent with other theoretical results as shown in Table~\ref{tab:mass}. Thus, we adopt two choices of mass as input to calculate the strong decay properties. First, assuming $h_1(1415)$ as $1^1P_1$ $s\bar{s}$ state with the predicted mass $M=1460$~MeV, and the corresponding total decay width is predicted to be $\Gamma=139$~MeV which is larger than the average experimental value $78\pm11$~MeV~\cite{ParticleDataGroup:2024cfk}. Second, using the experimental mass, $M=1409$ MeV, as input and the corresponding calculated width will be now $\Gamma=82$ MeV which is consistent with experimental data. It should be point out that with the assignment of $1^1P_1$ $s\bar{s}$ state, the $h_1(1415)$ state can only decay into $K\bar{K}^*$ channel.
\par
There are some different explanations for the $f_1(1420)$, such as candidate of the $1^3P_1$ $s\bar{s}$ state, hybrid state~\cite{Ishida:1989xh}, and $K\bar{K}^*$ molecule state~\cite{Longacre:1990uc}. Besides, in Ref.~\cite{Debastiani:2016xgg} by studying the production and decay of the $f_1(1285)$ into $\pi a_0(980)$ and $K^*\bar{K}$, the $f_1(1420)$ is not a
genuine resonance. The experimental information of $f_1(1420)$ has not been updated for a long time. The latest experimental mass and width are given by the L3 Collaboration in 2007 with $M=1434\pm5\pm5$~MeV,
and by the DELPHI Collaboration in 2003 with $M=1426\pm6$~MeV and $\Gamma=51\pm14$~MeV~\cite{DELPHI:2003bnm}. Considering $f_1(1420)$ as $1^3P_1$ $s\bar{s}$ state, we predict its mass $M=1464$~MeV, which is larger than the experimental value $1426$~MeV but is close to other theoretical results shown in Table~\ref{tab:mass}. The $K\bar{K}^*$ is the only dominant strong decay mode, which is the same $1^1P_1$ state. The width of $f_1(1420)$ is predicted to be $\Gamma=278$~MeV if theoretical mass is used as input, and $\Gamma=210$~MeV by using experimental mass, which are close to other theoretical results~\cite{Barnes:2002mu,Li:2020xzs}, but larger than the experimental data $\Gamma=54.5\pm2.6$~MeV. However, as discussed in Refs.~\cite{Barnes:2002mu,Li:2020xzs}, this discrepancy could be due to the nearby $K\bar{K}^*(892)$ threshold effects which can lead to strong suppression of the resonant width. 
\par
The $f_2^\prime(1525)$ is widely regarded as the $1^3P_2$ $s\bar{s}$ state~\cite{Barnes:2002mu,Godfrey:1985xj,Roberts:1997kq,Li:2020xzs}. Our predicted mass $1523$~MeV is consistent with experimental mass, but the predicted width is 30~MeV or 28~MeV which are less than the experimental value $86\pm5$~MeV. However, the branching fraction $\mathcal{B}(f_2^\prime(1525)\to K\bar{K})\approx70\%$ is close to the measured value $88\%$ from RPP~\cite{ParticleDataGroup:2024cfk}, and $70\%$ from Ref.~\cite{Li:2020xzs}. Besides, the branching fraction $\mathcal{B}(f_2^\prime(1525)\to\eta\eta)\approx10\%$ is also consistent with the results of Ref.~\cite{Li:2020xzs}. Furthermore, the ratio $R_{\eta\eta/K\bar{K}}={\Gamma_{\eta\eta}}/{\Gamma_{K\bar{K}}}=14\%$ is also close to $13\%$ from Ref.~\cite{Li:2020xzs} and $11.5\%$ from RPP~\cite{ParticleDataGroup:2024cfk}. It should be noticed that, although there is no experimental information available on the channel $f_2^\prime(1525)\to K\bar{K}^*(892)$, it is also an important decay mode. Its branching fraction is predicted to be $20\%$ in our model and $21\%$ in Ref.~\cite{Li:2020xzs}.

  It is noteworthy that the $f_2'(1525)$ and $f_0(1370)$  may be interpreted as molecular states formed through vector-vector interactions~\cite{Geng:2008gx,Dai:2013uua,Xie:2014gla,Nagahiro:2008bn,Abreu:2023yvf}. To investigate this possibility, we repeated our analysis by omitting the $f_2'(1525)$ from the input dataset. We found that the fitted parameters exhibit no substantial changes, and the resulting predictions remain consistent. Future precision measurements of the processes proposed in Refs.~\cite{Dai:2013uua,Xie:2014gla,Nagahiro:2008bn,Abreu:2023yvf} could help clarify the nature of these states.

\subsection{$2P$-wave $s\bar{s}$ states}

\begin{table*}[htpb]
\begin{center}
\caption{\label{tab:decay2} The decay widths (in MeV) of the $S$-wave and $P$-wave strangeonium. The data in \{\} means the width is calculated by using experimental initial state mass.}
\footnotesize
\begin{tabular}{cccccccccccccc}
\hline\hline
$n^{2S+1}L_J$    &$3^1S_0$ &$3^3S_1$ &$2^3P_0$ &$2^1P_1$ &$2^3P_1$ &$2^3P_2$ &$3^3P_0$ &$3^1P_1$ &$3^3P_1$ &$3^3P_2$\\\hline
$K\bar{K}$              &$-$  &11.2\{9.9\}  &51.8   &$-$          &$-$    &0.1    &24.8   &$-$          &$-$    &1.2      \\
$K^*\bar{K}$            &49.7 &43.3\{25.1\} &$-$    &58.7\{55.2\} &58.7   &15.8   &$-$    &39.4\{41.7\} &31.3   &19.2     \\
$K^*\bar{K}^*$          &12.0 &0.4\{6.5\}   &62.2   &38.1\{37.0\} &44.1   &21.1   &16.4   &13.4\{15.3\} &20.2   &14.4    \\
$K^*_0(1430)\bar{K}$    &54.3 &$-$          &$-$    &0.6          &0.1    &$-$    &$-$    &0.3\{0.5\}   &0.6    &$-$    \\
$K^*_0(1430)\bar{K}^*$  &$-$  &$-$          &$-$    &$-$          &$-$    &$-$    &$-$    &$-$          &$-$    &0.5      \\
$K_{1B}\bar{K}$         &$-$  &6.2\{9.0\}   &35.5   &0.8\{0.7\}   &0.2    &31.1   &2.5    &0.7\{0.7\}   &2.9    &11.3      \\
$K_{1B}\bar{K}^*$       &$-$  &$-$          &$-$    &$-$          &$-$    &$-$    &12.6   &24.9\{23.0\} &19.2   &6.1      \\
$K_{1A}\bar{K}$         &$-$  &17.2\{6.7\}  &3.9    &0.3\{0.1\}   &17.6   &0.2    &30.0   &0.8\{0.9\}   &11.2   &3.4      \\
$K_{1A}\bar{K}^*$       &$-$  &$-$          &$-$    &$-$          &$-$    &$-$    &$-$    &1.2\{6.0\}   &5.7    &2.2      \\
$K_2^*(1430)\bar{K}$    &20.2 &24.4\{4.1\}  &$-$    &3.6          &11.4   &7.4    &$-$    &29.4\{33.1\} &21.1   &14.3      \\
$K_2^*(1430)\bar{K}^*$  &$-$  &$-$          &$-$    &$-$          &$-$    &$-$    &$-$    &$-$          &$-$    &7.6      \\
$\eta\eta$              &$-$  &$-$          &1.4    &$-$          &$-$    &0.02   &0.4    &$-$          &$-$    &0.1  \\
$\eta\eta^\prime$       &$-$  &$-$          &0.1    &$-$          &$-$    &0.1    &0.03   &$-$          &$-$    &0.1   \\
$\eta^\prime\eta^\prime$&$-$  &$-$          &$-$    &$-$          &$-$    &0.01   &0.3    &$-$          &$-$    &0.1   \\
$\eta \phi$             &$-$  &1.0\{0.3\}   &$-$    &1.8\{1.5\}   &$-$    &$-$    &$-$    &0.9\{0.9\}   &$-$    &$-$     \\
$\eta^\prime \phi$      &$-$  &0.03\{0.01\} &$-$    &$-$          &$-$    &$-$    &$-$    &0.5\{0.5\}   &$-$    &$-$     \\
$K\bar{K}(1460)$        &$-$  &10.8\{1.2\}  &$-$    &$-$          &$-$    &$-$    &13.9   &$-$          &$-$    &1.0     \\
$\eta f_0(1370)$        &6.0  &$-$          &$-$    &$-$          &0.05   &$-$    &$-$    &$-$          &0.1    &$-$    \\
$\eta h_1(1415)$        &$-$  &1.1\{0.8\}   &$-$    &$-$          &$-$    &$-$    &$-$    &$-$          &$-$    &$-$  \\
$\eta f_1(1420)$        &4.9  &$-$          &$-$    &$-$          &$-$    &$-$    &$-$    &$-$          &0.1    &$-$   \\
$\eta f_2^\prime(1525)$ &$-$  &$-$          &$-$    &$-$          &$-$    &$-$    &$-$    &$-$          &0.2    &0.2   \\
Total                   &147.0&115.6\{63.5\}&154.9  &103.8\{94.5\} &132.1 &75.7   &101.0  &111.6\{122.7\}&112.5 &81.7      \\
Exp.                    &     &$148\pm16\pm66$ & &$149\pm12\pm23$ &   &     &     &$89\pm15\pm27$ &  &    \\
\hline\hline
\end{tabular}
\end{center}
\end{table*}

\begin{table*}[htpb]
\begin{center}
\caption{\label{tab:decay3} The decay widths (in MeV) of the $D$-wave strangeonium. The data in \{\} means the width is calculated by using experimental initial state mass.}
\footnotesize
\begin{tabular}{cccccccccccccc}
\hline\hline
$n^{2S+1}L_J$         &$1^3D_1$  &$1^1D_2$      &$1^3D_2$  &$1^3D_3$      &$2^3D_1$      &$2^1D_2$  &$2^3D_2$  &$2^3D_3$\\\hline
$K\bar{K}$               &50.1   &$-$           &$-$       &17.1\{17.2\}  &25.7\{24.2\}  &$-$       &$-$       &3.1       \\
$K^*\bar{K}$             &59.8   &92.9\{91.7\}  &123.3     &22.7\{22.9\}  &15.7\{12.4\}  &27.2      &38.9      &0.02     \\
$K^*\bar{K}^*$           &5.1    &24.2\{21.3\}  &14.3      &55.7\{56.7\}  &51.6\{40.5\}  &21.1      &26.5      &26.9       \\
$K^*_0(1430)\bar{K}$     &$-$    &$-$           &$-$       &$-$           &$-$           &6.4       &0.2       &$-$      \\
$K^*_0(1430)\bar{K}^*$   &$-$    &$-$           &$-$       &$-$           &$-$           &$-$       &$-$       &$-$       \\
$K_{1B}\bar{K}$          &57.0   &0.3\{0.2\}    &0.5       &2.2\{2.3\}    &20.9\{18.1\}  &0.1       &1.8       &10.1      \\
$K_{1B}\bar{K}^*$        &$-$    &$-$           &$-$       &$-$           &6.2\{7.1\}    &11.5      &12.9      &7.1        \\
$K_{1A}\bar{K}$          &$-$    &$-$           &$-$       &$-$           &37.4\{17.9\}  &0.4       &20.4      &3.2      \\
$K_{1A}\bar{K}^*$        &$-$    &$-$           &$-$       &$-$           &$-$           &$-$       &$-$       &$-$      \\
$K_2^*(1430)\bar{K}$     &$-$    &$-$           &$-$       &$-$           &22.2\{17.0\}  &40.3      &27.4      &13.4      \\
$K_2^*(1430)\bar{K}^*$   &$-$    &$-$           &$-$       &$-$           &$-$           &$-$       &$-$       &$-$      \\
$\eta\eta$               &$-$    &$-$           &$-$       &$-$           &$-$           &$-$       &$-$       &$-$      \\
$\eta\eta^\prime$        &$-$    &$-$           &$-$       &$-$           &$-$           &$-$       &$-$       &$-$     \\
$\eta \phi$              &2.9    &$-$           &12        &0.2\{0.2\}    &0.2\{0.1\}    &$-$       &1.0       &0.1      \\
$\eta^\prime \phi$       &$-$    &$-$           &$-$       &$-$           &0.1\{0.1\}    &$-$       &0.1       &0.1      \\
$K\bar{K}(1460)$         &$-$    &$-$           &$-$       &$-$           &23.9\{23.7\}  &$-$       &$-$       &3.7     \\
$\eta f_0(1370)$         &$-$    &$-$           &$-$       &$-$           &$-$           &0.4       &$-$       &$-$     \\
$\eta h_1(1415)$         &$-$    &$-$           &$-$       &$-$           &0.8\{0.5\}    &$-$       &0.2       &0.4     \\
$\eta f_1(1420)$         &$-$    &$-$           &$-$       &$-$           &$-$           &0.3       &$-$       &$-$    \\
$\eta f_2^\prime(1525)$  &$-$    &$-$           &$-$       &$-$           &$-$           &0.6       &$-$       &$-$     \\
Total                    &174.9  &117.3\{113.2\}&144.2     &97.9\{99.3\}  &204.7\{161.7\}&108.4     &129.3     &68.1      \\
Exp.                     &       &$225\pm14$&       &$87^{+28}_{-23}$&$106^{+24}_{-18}$  &          &          &    \\
\hline\hline
\end{tabular}
\end{center}
\end{table*}
We predicted the masses of the four $2P$-wave $s\bar{s}$ states, which are 1907~MeV, 1934~MeV, 1948~MeV, and 1969~MeV for $2^3P_0$, $2^1P_1$, $2^3P_1$, and $2^3P_2$, respectively. Due to the screening effects, our predicted results are smaller than the other predictions as shown in Table~\ref{tab:mass}. Comparing the mass information, one can find that the new observed state $h_1(1911)$($M=1911\pm6\pm14$ MeV) can be a good candidate as $h_1(2^1P_1)$ $s\bar{s}$ state. Besides, as shown in Table~\ref{tab:decay2}, the predicted total decay width is 112~MeV and 96~MeV, both of which are close to the measured data $\Gamma=149\pm12\pm23$~MeV. This further supports $h_1(2^1P_1)$ $s\bar{s}$ state assignment. The branching fractions of the dominantly channels are $\mathcal{B}(h_1(2^1P_1)\to K\bar{K}^*)\approx56\%$ and $\mathcal{B}(h_1(2^1P_1)\to K^*\bar{K}^*)\approx36\%$, respectively. 
  
\subsection{$3P$-wave $s\bar{s}$ states}
Recently, $X(2300)$ with mass of $2316\pm9\pm30$~MeV and width of $89\pm15\pm26$~MeV was firstly reported by the BESIII Collaboration. According to Table~\ref{tab:mass} and~\ref{tab:decay2}, $X(2300)$ could be a good candidate for $3^1P_1$ $s\bar{s}$ state with predicted mass 2301~MeV and width 112~MeV or 123~MeV. Its dominant decay modes are $K\bar{K}^*$, $K^*\bar{K}^*$, $K^*\bar{K}_{1B}$, and $K\bar{K}_2^*(1430)$ with branching fractions $35\%$, $12\%$, $22\%$, and $26\%$, respectively.
\par
For the other $3P$-wave $s\bar{s}$ states, our predicted masses and widths are $M(3^3P_0)=2290$~MeV, $M(3^3P_1)=2314$~MeV, $M(3^3P_2)=2329$~MeV, $\Gamma(3^3P_0)=102$~MeV, $\Gamma(3^3P_1)=113$~MeV, and $\Gamma(3^3P_2)=82$~MeV. The predicted masses are smaller than the predictions of the other models due to the screening effects. As shown in Table~\ref{tab:decay2}, the main decay modes of $3^3P_0$ $s\bar{s}$ state are $K\bar{K}~(25\%)$, $K^*\bar{K}^*~(16\%)$, $K^*\bar{K}_{1B}~(13\%)$, $K\bar{K}_{1A}~(29\%)$, and $K\bar{K}(1460)~(14\%)$, and the ones of $3^3P_1$ $s\bar{s}$ state are $K\bar{K}^*~(27\%)$, $K^*\bar{K}^*~(18\%)$, $K^*\bar{K}_{1B}~(17\%)$, $K\bar{K}_{1A}~(10\%)$, and $K\bar{K}_2^*(1430)~(19\%)$. The $3^3P_2$ $s\bar{s}$ state can mainly decay into $K\bar{K}^*~(23\%)$, $K^*\bar{K}^*~(17\%)$, $K\bar{K}_{1B}~(13\%)$, and $K\bar{K}_2^*(1430)~(17\%)$ final states.

\subsection{$1D$-wave $s\bar{s}$ states}
The $1^3D_1$ $s\bar{s}$ state has not been established in experiments yet. The predicted  mass within the MGI model is 1832~MeV, which is comparable with mass range $1750\sim 1880$~MeV of the various quark models as shown in Table~\ref{tab:mass}. With the wave functions obtained using the MGI model, the strong decay width is predicted to be 178~MeV and the dominant decay modes are $K\bar{K}$, $K^*\bar{K}$, $K^*\bar{K}^*$, $K_{1B}\bar{K}$ and $\eta \phi$. 
\par
The $\eta_2(1870)$ was first observed by the Crystal Ball Collaboration in the process $e^+e^-\to e^+e^-\eta \pi^0\pi^0$ with mass $1881\pm32\pm40$ MeV and width $221\pm92\pm44$~MeV~\cite{CrystalBall:1991zkb}. Based on the results of multiple experimental groups, the current average experimental values of mass and width for $\eta_2(1870)$ with quantum numbers $I^G(J^{PC})=0^+(2^{-+})$ is $M=1842\pm8$~MeV and $\Gamma=225\pm14$~MeV~~\cite{ParticleDataGroup:2024cfk}, which are consistent with our prediction for $1^1D_2$ state, as discussed in Ref.~\cite{Li:2009rka}. Our calculations show that this state mainly decays into $K^*\bar{K}$ and $K^*\bar{K}^*$, which should be helpful to search for this state in experiments.
\par
There is no candidate for $1^3D_2$ state by now. The predicted mass of $1^3D_2$ state is  1857~MeV,  close to the one of $1^1D_2$ state. However, as shown in Table~\ref{tab:decay3}, the $1^3D_2$ state could decay into $\eta\phi$ channel that is forbidden for $1^1D_2$ due to the charge-parity conservation, which could be used to distinguish these two states in future.
\par
The $\phi_3(1850)$ was first found in the process $K^-p\to K\bar{K}\Lambda$ of a high statistics bubble chamber experiment~\cite{BIRMINGHAM-CERN-GLASGOW-MICHIGANSTATE-PARIS:1981fry}. The predicted mass, 1853~MeV, and width, 98~MeV, for $1^3D_3$ state are in consistent with experimental data of $M=1854\pm7$~MeV and $\Gamma=87_{-23}^{+28}$~MeV. 
Besides, the decay ratio predicted in this work is
\begin{equation}
\frac{\Gamma(K\bar{K}^*)}{\Gamma(K\bar{K})} = 1.35,
\end{equation} 
which is also in consistent with the LASS experimental data $0.55^{+0.85}_{-0.45}$ of Ref.~\cite{Aston:1988rf}.

\subsection{$2D$-wave $s\bar{s}$ states}
For the $2D$-wave states, combining our and other theoretical results as listed in Table~\ref{tab:mass}, we can estimate the mass ranges of the four $2D$-wave states as $2220\sim 2342$ MeV, $2220\sim 2340$ MeV, $2229\sim 2348$ MeV and $2223\sim 2360$ MeV.
\par
The $\phi(2170)$ was found in 2006 in the process $e^+e^-\to K^+K^-\pi\pi\gamma$ by BABAR Collaboration with fitted mass $2175\pm10\pm15$~MeV and width $58\pm16\pm20$~MeV~\cite{BaBar:2006gsq}. Subsequently, a large amount of experimental works have confirmed the existence of this state. The latest experimental results reported by BESIII in 2023 indicates that its mass is $2178\pm20\pm5$~MeV and width is $140\pm36\pm16$~ MeV~\cite{BESIII:2021aet}, and its internal structure are still in debate~\cite{Zhao:2019syt,Jiang:2023atq}. Now the quantum numbers of this state are determined to be $I^G(J^{PC})=0^-(1^{--})$. As shown in Table~II and Table~V, our predicted mass and width for the $2^3D_1$ $s\bar{s}$ state are in agreement with the experimental data.
Our results show that this state can decay into the modes $K\bar{K}$($13\%$), $K^*\bar{K}$($8\%$), $K^*\bar{K^*}$($25\%$), $K_{1B}\bar{K}$($10\%$), $K_{1B}\bar{K}^*$($3\%$), $K_{1A}\bar{K}$($18\%$), $K_2^*(1430)\bar{K}$($11\%$), $K\bar{K}(1460)$($12\%$) and $\eta h_1(1415)$($1\%$), while the $K^*\bar K^*$ mode is highly suppressed and the decay to $\phi\eta$ and 
$\phi\eta^\prime$ have been observed, according to the recent BESIII results. Indeed, many studies show that the $\phi(2170)$ have a molecular nature~\cite{Malabarba:2023zez,Malabarba:2020grf,MartinezTorres:2008gy,Alvarez-Ruso:2009vkn}.

\par
We have also calculated the masses the decay properties of the $2^1D_2$, $2^3D_2$, and $2^3D_3$ states, which will be helpful in the future experiments. It is shown that $2^1D_2$ state can decay into $K^*\bar{K}$($25\%$), $K^*\bar{K}^*$($19\%$), $K_0^*(1430)\bar{K}$~($5\%$), $K_{1B}\bar{K}^*$($10\%$), $K_2^*(1430)\bar{K}$($36\%$), $\eta f_0(1370)$($1\%$), $\eta f_1(1420)$($1\%$) and $\eta f_2^\prime(1525)$($1\%$). The decay modes for $2^3D_2$ state are $K^*\bar{K}$($30\%$), $K^*\bar{K}^*$($21\%$), $K_{1B}\bar{K}$($2\%$), $K_{1B}\bar{K}^*$($10\%$), $K_{1A}\bar{K}$($15\%$), $K_2^*(1430)\bar{K}$($21\%$) and $\eta \phi$($2\%$). For the $2^3D_3$ state, it can decay into $K\bar{K}$($4\%$), $K^*\bar{K}^*$($39\%$), $K_{1B}\bar{K}$($14\%$), $K_{1B}\bar{K}^*$($10\%$), $K_{1A}\bar{K}$($4\%$), $K_2^*(1430)\bar{K}$($19\%$), $K\bar{K}(1460)$($6\%$), and $\eta h_1(1415)$($1\%$). Comparing the proportions of each channel, the decay channel $K\bar{K}$ plays an important role in distinguishing states  $2^3D_1$, $2^3D_3$ from $2^1D_2$, $2^3D_2$, because the last two states are forbidden to decay into $K\bar{K}$. Besides, the proportion of $K\bar{K}$($13\%$) for state $2^3D_1$ is larger than the one ($4\%$) for the $2^3D_1$ state. For the $2^1D_2$ and $2^3D_2$, our results show that the $2^3D_2$ state can easily decay into the mode $K_{1A}\bar{K}$ ($15\%$), which is forbidden for the $2^1D_2$ state.

\section{Summary}
\label{sec:summary}
Inspired by new observed states $h_1(1911)$ and $X(2300)$, we have systematically investigated strangeonium mesons within the modified GI model and the $^3P_0$ model. In our calculations, meson space wave functions obtained using the MGI potential model are adopted to calculate the strong decay widths. 
\par
Our calculations indicate that $h_1(1911)$ can be regarded as a good candidate for $2^1P_1$ $s\bar{s}$ state with predicted mass 1934~MeV and total width 106~MeV, which are in consistent with measured data $M=1911\pm6\pm14$~MeV and $\Gamma=149\pm12\pm23$~MeV.  The branching fractions of the dominant decay modes are $\mathcal{B}(h_1(2^1P_1)\to KK^*)\approx56\%$ and $\mathcal{B}(h_1(2^1P_1)\to K^*K^*)\approx36\%$. The $X(2300)$ with mass $2316\pm9\pm30$~MeV and width $89\pm15\pm26$~MeV can be well interpreted as $3^1P_1$ $s\bar{s}$ state. The experimental information are in good agreement with our calculation of the mass 2301~MeV and  width 112~MeV. Furthermore, the predicted dominant decay modes are $K\bar{K}^*$, $K^*\bar{K}^*$, $K^*\bar{K}_{1B}$, and $K\bar{K}_2^*(1430)$.
\par
Furthermore, we have also calculate the spectrum and strong decay properties of $S$-wave, $P$-wave, $D$-wave  $s\bar{s}$ mesons. The possible assignments of $X(2000)$, $\eta_2(1870)$ as $3^3S_1$ and $1^1D_2$ are discussed. For the $2D$-wave state, our results support $2^3D_1$ assignment for $\phi(2170)$, which is consistent with the conclusion of Refs.~\cite{Pang:2019ttv,Li:2020xzs}.

\section{Acknowledgements}

This work is supported by the National Key R\&D Program of China (No. 2024YFE0105200), the Natural Science Foundation of Henan under Grant No. 232300421140 and No. 222300420554, the National Natural Science Foundation of China under Grant No. 12475086, No. 12447208, and No. 12192263.

\bibliographystyle{unsrt}
\bibliography{cite}  

\begin{thebibliography}{10}

\bibitem{Liu:2015zqa}
Pei-Lian Liu, Shuang-Shi Fang, and Xin-Chou Lou.
\newblock {Strange Quarkonium States at BESIII}.
\newblock {\em Chin. Phys. C}, 39(8):082001, 2015.

\bibitem{Ketzer:2019wmd}
Bernhard Ketzer, Boris Grube, and Dmitry Ryabchikov.
\newblock {Light-Meson Spectroscopy with COMPASS}.
\newblock {\em Prog. Part. Nucl. Phys.}, 113:103755, 2020.

\bibitem{Brambilla:2019esw}
Nora Brambilla, Simon Eidelman, Christoph Hanhart, Alexey Nefediev, Cheng-Ping Shen, Christopher~E. Thomas, Antonio Vairo, and Chang-Zheng Yuan.
\newblock {The $XYZ$ states: experimental and theoretical status and perspectives}.
\newblock {\em Phys. Rept.}, 873:1--154, 2020.

\bibitem{vanBeveren:2020eis}
Eef van Beveren and George Rupp.
\newblock {Modern meson spectroscopy: the fundamental role of unitarity}.
\newblock {\em Prog. Part. Nucl. Phys.}, 117:103845, 2021.

\bibitem{Li:2020xzs}
Qi~Li, Long-Cheng Gui, Ming-Sheng Liu, Qi-Fang L\"u, and Xian-Hui Zhong.
\newblock {Mass spectrum and strong decays of strangeonium in a constituent quark model}.
\newblock {\em Chin. Phys. C}, 45(2):023116, 2021.

\bibitem{Godfrey:1985xj}
S.~Godfrey and Nathan Isgur.
\newblock {Mesons in a Relativized Quark Model with Chromodynamics}.
\newblock {\em Phys. Rev. D}, 32:189--231, 1985.

\bibitem{Burakovsky:1997dd}
L.~Burakovsky and J.~Terrance Goldman.
\newblock {Constraint on axial - vector meson mixing angle from nonrelativistic constituent quark model}.
\newblock {\em Phys. Rev. D}, 56:R1368--R1372, 1997.

\bibitem{Pang:2019ttv}
Cheng-Qun Pang.
\newblock {Excited states of $\phi$ meson}.
\newblock {\em Phys. Rev. D}, 99(7):074015, 2019.

\bibitem{Gao:2019jme}
Ya~Gao, Yateng Zhang, Bo~Zheng, Zhen-Hua Zhang, Wenbiao Yan, and Xiaohua Li.
\newblock {Evaluation of the $\theta_K$ and the Mixing angle $\theta _{h_1}$}.
\newblock {\em arXiv: 1911.06967}, 2019.

\bibitem{Cheng:2011pb}
Hai-Yang Cheng.
\newblock {Revisiting Axial-Vector Meson Mixing}.
\newblock {\em Phys. Lett. B}, 707:116--120, 2012.

\bibitem{Burakovsky:1997ci}
L.~Burakovsky and J.~Terrance Goldman.
\newblock {Towards resolution of the enigmas of $P$ wave meson spectroscopy}.
\newblock {\em Phys. Rev. D}, 57:2879--2888, 1998.

\bibitem{Li:2005eq}
De-Min Li, Bing Ma, and Hong Yu.
\newblock {Regarding the axial-vector mesons}.
\newblock {\em Eur. Phys. J. A}, 26:141--145, 2005.

\bibitem{BESIII:2023zwx}
M.~Ablikim et~al.
\newblock {Study of the decay $J/\psi \to \phi \pi^{0}\eta$}.
\newblock 11 2023.

\bibitem{BESIII:2018zbm}
Medina Ablikim et~al.
\newblock {Observation and study of the decay $J/\psi\rightarrow\phi\eta\eta'$}.
\newblock {\em Phys. Rev. D}, 99(11):112008, 2019.

\bibitem{BESIII:2024nhv}
Medina Ablikim et~al.
\newblock {Observation of an axial-vector state in the study of $\psi(3686) \to \phi \eta \eta'$ decay}.
\newblock 10 2024.

\bibitem{Brown:1979ya}
Lowell~S. Brown and William~I. Weisberger.
\newblock {Remarks on the Static Potential in Quantum Chromodynamics}.
\newblock {\em Phys. Rev. D}, 20:3239, 1979.

\bibitem{Eichten:1980mw}
E.~Eichten and F.~Feinberg.
\newblock {Spin Dependent Forces in QCD}.
\newblock {\em Phys. Rev. D}, 23:2724, 1981.

\bibitem{Gupta:1981pd}
S.~N. Gupta and S.~F. Radford.
\newblock {Quark Quark and Quark - Anti-quark Potentials}.
\newblock {\em Phys. Rev. D}, 24:2309--2323, 1981.

\bibitem{Pantaleone:1985uf}
James~T. Pantaleone, S.~H.~Henry Tye, and Yee~Jack Ng.
\newblock {Spin Splittings in Heavy Quarkonia}.
\newblock {\em Phys. Rev. D}, 33:777, 1986.

\bibitem{Lakhina:2006fy}
Olga Lakhina and Eric~S. Swanson.
\newblock {A Canonical $D_s(2317)$?}
\newblock {\em Phys. Lett. B}, 650:159--165, 2007.

\bibitem{Lu:2016mbb}
Yu~Lu, Muhammad~Naeem Anwar, and Bing-Song Zou.
\newblock {Coupled-Channel Effects for the Bottomonium with Realistic Wave Functions}.
\newblock {\em Phys. Rev. D}, 94(3):034021, 2016.

\bibitem{Ferretti:2012zz}
J.~Ferretti, G.~Galata, E.~Santopinto, and A.~Vassallo.
\newblock {Bottomonium self-energies due to the coupling to the meson-meson continuum}.
\newblock {\em Phys. Rev. C}, 86:015204, 2012.

\bibitem{Ferretti:2013faa}
J.~Ferretti, G.~Galat\`a, and E.~Santopinto.
\newblock {Interpretation of the X(3872) as a charmonium state plus an extra component due to the coupling to the meson-meson continuum}.
\newblock {\em Phys. Rev. C}, 88(1):015207, 2013.

\bibitem{Ortega:2016mms}
Pablo~G. Ortega, Jorge Segovia, David~R. Entem, and Francisco Fernandez.
\newblock {Molecular components in $P$-wave charmed-strange mesons}.
\newblock {\em Phys. Rev. D}, 94(7):074037, 2016.

\bibitem{Hao:2022vwt}
Wei Hao, Yu~Lu, and Bing-Song Zou.
\newblock {Coupled channel effects for the charmed-strange mesons}.
\newblock {\em Phys. Rev. D}, 106(7):074014, 2022.

\bibitem{Yang:2023tvc}
Jing-Jing Yang, Wei Hao, Xiaoyu Wang, De-Min Li, Yu-Xiao Li, and En~Wang.
\newblock {The mass spectrum and strong decay properties of the charmed-strange mesons within Godfrey\textendash{}Isgur model considering the coupled-channel effects}.
\newblock {\em Eur. Phys. J. C}, 83(12):1098, 2023.

\bibitem{Hao:2024nqb}
Wei Hao and Ruilin Zhu.
\newblock {Beauty-charm meson family with coupled channel effects and their strong decays}.
\newblock {\em Chin. Phys. C}, 48(12):123101, 2024.

\bibitem{Hao:2024ptu}
Wei Hao, M.~Atif Sultan, and En~Wang.
\newblock {Spectrum and decay properties of the charmed mesons involving the coupled channel effects}.
\newblock {\em arXiv: 2411.02976}, 2024.

\bibitem{Laermann:1986pu}
E.~Laermann, F.~Langhammer, I.~Schmitt, and P.~M. Zerwas.
\newblock {The Interquark Potential: SU(2) Color Gauge Theory With Fermions}.
\newblock {\em Phys. Lett. B}, 173:437--442, 1986.

\bibitem{Born:1989iv}
K.~D. Born, E.~Laermann, N.~Pirch, T.~F. Walsh, and P.~M. Zerwas.
\newblock {Hadron Properties in Lattice {QCD} With Dynamical Fermions}.
\newblock {\em Phys. Rev. D}, 40:1653--1663, 1989.

\bibitem{Armoni:2008jy}
Adi Armoni.
\newblock {Beyond The Quenched (or Probe Brane) Approximation in Lattice (or Holographic) QCD}.
\newblock {\em Phys. Rev. D}, 78:065017, 2008.

\bibitem{Li:2009ad}
Bai-Qing Li, Ce~Meng, and Kuang-Ta Chao.
\newblock {Coupled-Channel and Screening Effects in Charmonium Spectrum}.
\newblock {\em Phys. Rev. D}, 80:014012, 2009.

\bibitem{Hao:2019fjg}
Wei Hao, Guan-Ying Wang, En~Wang, Guan-Nan Li, and De-Min Li.
\newblock {Canonical interpretation of the $X(4140)$ state within the $^3P_0$ model}.
\newblock {\em Eur. Phys. J. C}, 80(7):626, 2020.

\bibitem{Feng:2022esz}
Xue-Chao Feng, Wei Hao, and li-Juan Liu.
\newblock {The assignments of the bottom mesons within the screened potential model and $^3P_0$ model}.
\newblock {\em Int. J. Mod. Phys. E}, 31(07):2250066, 2022.

\bibitem{Hao:2022ibj}
Wei Hao, Yu~Lu, and En~Wang.
\newblock {The assignments of the $B_s$ mesons within the screened potential model and $^3P_0$ model}.
\newblock {\em Eur. Phys. J. C}, 83(6):520, 2023.

\bibitem{Song:2015fha}
Qin-Tao Song, Dian-Yong Chen, Xiang Liu, and Takayuki Matsuki.
\newblock {Higher radial and orbital excitations in the charmed meson family}.
\newblock {\em Phys. Rev. D}, 92(7):074011, 2015.

\bibitem{Song:2015nia}
Qin-Tao Song, Dian-Yong Chen, Xiang Liu, and Takayuki Matsuki.
\newblock {Charmed-strange mesons revisited: mass spectra and strong decays}.
\newblock {\em Phys. Rev. D}, 91:054031, 2015.

\bibitem{Wang:2019mhs}
Jun-Zhang Wang, Dian-Yong Chen, Xiang Liu, and Takayuki Matsuki.
\newblock {Constructing $J/\psi$ family with updated data of charmoniumlike $Y$ states}.
\newblock {\em Phys. Rev. D}, 99(11):114003, 2019.

\bibitem{Wang:2018rjg}
Jun-Zhang Wang, Zhi-Feng Sun, Xiang Liu, and Takayuki Matsuki.
\newblock {Higher bottomonium zoo}.
\newblock {\em Eur. Phys. J. C}, 78(11):915, 2018.

\bibitem{Wang:2024lba}
Ya-Rong Wang, Xiao-Hai Liu, Cheng-Qun Pang, and Hao Chen.
\newblock {New states X(1910) and X(2300) and higher light excited JPC=1+- mesons}.
\newblock {\em Phys. Rev. D}, 111(5):054005, 2025.

\bibitem{Li:2021qgz}
Zheng-Ya Li, De-Min Li, En~Wang, Wen-Cheng Yan, and Qin-Tao Song.
\newblock {Assignments of the Y(2040), \ensuremath{\rho}(1900), and \ensuremath{\rho}(2150) in the quark model}.
\newblock {\em Phys. Rev. D}, 104(3):034013, 2021.

\bibitem{Feng:2022hwq}
Xue-Chao Feng, Zheng-Ya Li, De-Min Li, Qin-Tao Song, En~Wang, and Wen-Cheng Yan.
\newblock {Mass spectra and decay properties of the higher excited \ensuremath{\rho} mesons}.
\newblock {\em Phys. Rev. D}, 106(7):076012, 2022.

\bibitem{Micu:1968mk}
L.~Micu.
\newblock {Decay rates of meson resonances in a quark model}.
\newblock {\em Nucl. Phys. B}, 10:521--526, 1969.

\bibitem{LeYaouanc:1972vsx}
A.~Le~Yaouanc, L.~Oliver, O.~Pene, and J.~C. Raynal.
\newblock {Naive quark pair creation model of strong interaction vertices}.
\newblock {\em Phys. Rev. D}, 8:2223--2234, 1973.

\bibitem{LeYaouanc:1973ldf}
A.~Le~Yaouanc, L.~Oliver, O.~Pene, and J.~C. Raynal.
\newblock {Naive quark pair creation model and baryon decays}.
\newblock {\em Phys. Rev. D}, 9:1415--1419, 1974.

\bibitem{Li:2009rka}
De-Min Li and En~Wang.
\newblock {Canonical interpretation of the $\eta_2(1870)$}.
\newblock {\em Eur. Phys. J. C}, 63:297--304, 2009.

\bibitem{Pan:2016bac}
Ting-Ting Pan, Qi-Fang L\"u, En~Wang, and De-Min Li.
\newblock {Strong decays of the $X(2500)$ newly observed by the BESIII Collaboration}.
\newblock {\em Phys. Rev. D}, 94(5):054030, 2016.

\bibitem{Lu:2016bbk}
Qi-Fang L\"u, Ting-Ting Pan, Yan-Yan Wang, En~Wang, and De-Min Li.
\newblock {Excited bottom and bottom-strange mesons in the quark model}.
\newblock {\em Phys. Rev. D}, 94(7):074012, 2016.

\bibitem{Wang:2017pxm}
Guan-Ying Wang, Shi-Chen Xue, Guan-Nan Li, En~Wang, and De-Min Li.
\newblock {Strong decays of the higher isovector scalar mesons}.
\newblock {\em Phys. Rev. D}, 97(3):034030, 2018.

\bibitem{Xue:2018jvi}
Shi-Chen Xue, Guan-Ying Wang, Guan-Nan Li, En~Wang, and De-Min Li.
\newblock {The possible members of the $5^1S_0$ meson nonet}.
\newblock {\em Eur. Phys. J. C}, 78(6):479, 2018.

\bibitem{Ferretti:2013vua}
J.~Ferretti and E.~Santopinto.
\newblock {Higher mass bottomonia}.
\newblock {\em Phys. Rev. D}, 90(9):094022, 2014.

\bibitem{Jacob:1959at}
M.~Jacob and G.~C. Wick.
\newblock {On the General Theory of Collisions for Particles with Spin}.
\newblock {\em Annals Phys.}, 7:404--428, 1959.

\bibitem{Hayne:1981zy}
Cameron Hayne and Nathan Isgur.
\newblock {Beyond the Wave Function at the Origin: Some Momentum Dependent Effects in the Nonrelativistic Quark Model}.
\newblock {\em Phys. Rev. D}, 25:1944, 1982.

\bibitem{ParticleDataGroup:2024cfk}
S.~Navas et~al.
\newblock {Review of particle physics}.
\newblock {\em Phys. Rev. D}, 110(3):030001, 2024.

\bibitem{Xiao:2019qhl}
Li-Ye Xiao, Xin-Zhen Weng, Xian-Hui Zhong, and Shi-Lin Zhu.
\newblock {A possible explanation of the threshold enhancement in the process $e^+e^-\rightarrow \Lambda\bar{\Lambda}$}.
\newblock {\em Chin. Phys. C}, 43(11):113105, 2019.

\bibitem{Ebert:2009ub}
D.~Ebert, R.~N. Faustov, and V.~O. Galkin.
\newblock {Mass spectra and Regge trajectories of light mesons in the relativistic quark model}.
\newblock {\em Phys. Rev. D}, 79:114029, 2009.

\bibitem{Ishida:1986vn}
Shin Ishida and Kenji Yamada.
\newblock {Light Quark Meson Spectrum in the Covariant Oscillator Quark Model With One Gluon Exchange Effects}.
\newblock {\em Phys. Rev. D}, 35:265, 1987.

\bibitem{Vijande:2004he}
J.~Vijande, F.~Fernandez, and A.~Valcarce.
\newblock {Constituent quark model study of the meson spectra}.
\newblock {\em J. Phys. G}, 31:481, 2005.

\bibitem{Godfrey:2016nwn}
Stephen Godfrey, K.~Moats, and E.~S. Swanson.
\newblock {$B$ and $B_s$ Meson Spectroscopy}.
\newblock {\em Phys. Rev. D}, 94(5):054025, 2016.

\bibitem{Bertanza:1962zz}
L.~Bertanza et~al.
\newblock {Possible Resonances in the $\Xi\pi$ and $K\bar{K}$ Systems}.
\newblock {\em Phys. Rev. Lett.}, 9:180--183, 1962.

\bibitem{Mane:1982si}
F.~Mane, D.~Bisello, J.~C. Bizot, J.~Buon, A.~Cordier, and B.~Delcourt.
\newblock {Study of $e^+ e^- \to K^0_S K^\pm \pi^\mp$ in the 1.4-{GeV} to 2.18-{GeV} Energy Range: A New Observation of an Isoscalar Vector Meson $\phi^\prime$ (1.65-{GeV})}.
\newblock {\em Phys. Lett. B}, 112:178--182, 1982.

\bibitem{BESIII:2018ubj}
M.~Ablikim et~al.
\newblock {Amplitude analysis of the $K_{S}K_{S}$ system produced in radiative $J/\psi$ decays}.
\newblock {\em Phys. Rev. D}, 98(7):072003, 2018.

\bibitem{CrystalBarrel:2019zqh}
M.~Albrecht et~al.
\newblock {Coupled channel analysis of ${\bar{p}p}\,\rightarrow \,\pi ^0\pi ^0\eta $, ${\pi ^0\eta \eta }$ and ${K^+K^-\pi ^0}$ at 900 MeV/c and of ${\pi \pi }$-scattering data}.
\newblock {\em Eur. Phys. J. C}, 80(5):453, 2020.

\bibitem{Sarantsev:2021ein}
A.~V. Sarantsev, I.~Denisenko, U.~Thoma, and E.~Klempt.
\newblock {Scalar isoscalar mesons and the scalar glueball from radiative $J/\psi$ decays}.
\newblock {\em Phys. Lett. B}, 816:136227, 2021.

\bibitem{BESIII:2018ede}
Medina Ablikim et~al.
\newblock {Observation of $h_1(1380)$ in the $J/\psi \to \eta^{\prime} K\bar K \pi$ decay}.
\newblock {\em Phys. Rev. D}, 98(7):072005, 2018.

\bibitem{BESIII:2022zel}
M.~Ablikim et~al.
\newblock {Partial wave analysis of $J/\psi \rightarrow \gamma \eta' \eta'$}.
\newblock {\em Phys. Rev. D}, 105(7):072002, 2022.

\bibitem{Jiang:2019ijx}
Sheng-Juan Jiang, S.~Sakai, Wei-Hong Liang, and E.~Oset.
\newblock {The $\chi_{cJ}$ decay to $\phi K^* \bar K, \phi h_1(1380)$ testing the nature of axial vector meson resonances}.
\newblock {\em Phys. Lett. B}, 797:134831, 2019.

\bibitem{Roca:2005nm}
L.~Roca, E.~Oset, and J.~Singh.
\newblock {Low lying axial-vector mesons as dynamically generated resonances}.
\newblock {\em Phys. Rev. D}, 72:014002, 2005.

\bibitem{Ishida:1989xh}
Shin Ishida, Masuho Oda, Haruhiko Sawazaki, and Kenji Yamada.
\newblock {Is the $f_1(1420)$ Our First Hybrid Meson?}
\newblock {\em Prog. Theor. Phys.}, 82:119, 1989.

\bibitem{Longacre:1990uc}
Ronald~S. Longacre.
\newblock {The $E(1420)$ Meson as a $K \bar{K} \pi$ Molecule}.
\newblock {\em Phys. Rev. D}, 42:874--883, 1990.

\bibitem{Debastiani:2016xgg}
V.~R. Debastiani, F.~Aceti, Wei-Hong Liang, and E.~Oset.
\newblock {Revising the $f_1(1420)$ resonance}.
\newblock {\em Phys. Rev. D}, 95(3):034015, 2017.

\bibitem{DELPHI:2003bnm}
J.~Abdallah et~al.
\newblock {Measurement of inclusive $f_1(1285)$ and $f_1(1420)$ production in Z decays with the DELPHI detector}.
\newblock {\em Phys. Lett. B}, 569:129--139, 2003.

\bibitem{Barnes:2002mu}
T.~Barnes, N.~Black, and P.~R. Page.
\newblock {Strong decays of strange quarkonia}.
\newblock {\em Phys. Rev. D}, 68:054014, 2003.

\bibitem{Roberts:1997kq}
W.~Roberts and B.~Silvestre-Brac.
\newblock {Meson decays in a quark model}.
\newblock {\em Phys. Rev. D}, 57:1694--1702, 1998.

\bibitem{Geng:2008gx}
L.~S. Geng and E.~Oset.
\newblock {Vector meson-vector meson interaction in a hidden gauge unitary approach}.
\newblock {\em Phys. Rev. D}, 79:074009, 2009.

\bibitem{Dai:2013uua}
Lianrong Dai and Eulogio Oset.
\newblock {Tests on the molecular structure of $f_2(1270)$, $f'_2(1525)$ from $\psi (nS)$ and $\Upsilon (nS)$ decays}.
\newblock {\em Eur. Phys. J. A}, 49:130, 2013.

\bibitem{Xie:2014gla}
Ju-Jun Xie and E.~Oset.
\newblock {$\bar{B}^0$ and $\bar{B}^0_s$ decays into $J/\psi$ and $f_0(1370),~f_0(1710),~f_2(1270),~f'_2(1525),~K^*_2(1430)$}.
\newblock {\em Phys. Rev. D}, 90(9):094006, 2014.

\bibitem{Nagahiro:2008bn}
H.~Nagahiro, L.~Roca, E.~Oset, and B.~S. Zou.
\newblock {Role of meson loops in the f0 (1370), f0 (1500), and f0 (1710) decays into V gamma}.
\newblock {\em Phys. Rev. D}, 78:014012, 2008.

\bibitem{Abreu:2023yvf}
Luciano~M. Abreu, Lianrong Dai, and Eulogio Oset.
\newblock {J/\ensuremath{\psi} decay to \ensuremath{\phi},\ensuremath{\omega},K\textasteriskcentered{}0 plus f0(1370), f0(1710), K0\textasteriskcentered{}(1430), f2(1270), f2'(1525) and K2\textasteriskcentered{}(1430): Role of the D-wave for tensor production}.
\newblock {\em Phys. Lett. B}, 843:137999, 2023.

\bibitem{CrystalBall:1991zkb}
K.~Karch et~al.
\newblock {Analysis of the $\eta \pi^0 \pi^0$ final state in photon-photon collisions}.
\newblock {\em Z. Phys. C}, 54:33--44, 1992.

\bibitem{BIRMINGHAM-CERN-GLASGOW-MICHIGANSTATE-PARIS:1981fry}
S.~Al-Harran et~al.
\newblock {Observation of a $K \bar{K}$ Enhancement at 1.85-{GeV} in the Reaction $K^- P \to K \bar{K} \Lambda$ at 8.25-{GeV}/c}.
\newblock {\em Phys. Lett. B}, 101:357--360, 1981.

\bibitem{Aston:1988rf}
D.~Aston et~al.
\newblock {Spin Parity Determination of the $\phi_J(1850)$ From $K^- p$ Interactions at 11-{GeV}/$c$}.
\newblock {\em Phys. Lett. B}, 208:324--330, 1988.

\bibitem{BaBar:2006gsq}
Bernard Aubert et~al.
\newblock {A Structure at 2175-MeV in $e^{+} e^{-} \to \phi f_0(980)$ Observed via Initial-State Radiation}.
\newblock {\em Phys. Rev. D}, 74:091103, 2006.

\bibitem{BESIII:2021aet}
Medina Ablikim et~al.
\newblock {Measurement of $e^+e^-\to \phi\pi^+\pi^-$ cross sections at center-of-mass energies from 2.00 to 3.08~GeV}.
\newblock {\em Phys. Rev. D}, 108(3):032011, 2023.

\bibitem{Zhao:2019syt}
Chen-Guang Zhao, Guan-Ying Wang, Guan-Nan Li, En~Wang, and De-Min Li.
\newblock {$\phi(2170)$(2170) production in the process $\gamma p\to \eta \phi p$}.
\newblock {\em Phys. Rev. D}, 99(11):114014, 2019.

\bibitem{Jiang:2023atq}
Yi-Wei Jiang, Wei-Han Tan, Hua-Xing Chen, and Er-Liang Cui.
\newblock {Strong Decays of the $\phi$(2170) as a Fully Strange Tetraquark State}.
\newblock {\em Symmetry}, 16(8):1021, 2024.

\bibitem{Malabarba:2023zez}
Brenda~B. Malabarba, K.~P. Khemchandani, and A.~Martinez~Torres.
\newblock {\ensuremath{\phi}(2170) decaying to \ensuremath{\phi}\ensuremath{\eta} and \ensuremath{\phi}\ensuremath{\eta}'}.
\newblock {\em Phys. Rev. D}, 108(3):036010, 2023.

\bibitem{Malabarba:2020grf}
Brenda~B. Malabarba, Xiu-Lei Ren, K.~P. Khemchandani, and A.~Martinez~Torres.
\newblock {Partial decay widths of $\phi(2170)$ to kaonic resonances}.
\newblock {\em Phys. Rev. D}, 103(1):016018, 2021.

\bibitem{MartinezTorres:2008gy}
A.~Martinez~Torres, K.~P. Khemchandani, L.~S. Geng, M.~Napsuciale, and E.~Oset.
\newblock {The X(2175) as a resonant state of the phi K anti-K system}.
\newblock {\em Phys. Rev. D}, 78:074031, 2008.

\bibitem{Alvarez-Ruso:2009vkn}
L.~Alvarez-Ruso, J.~A. Oller, and J.~M. Alarcon.
\newblock {On the phi(1020) f0(980) S-wave scattering and the Y(2175) resonance}.
\newblock {\em Phys. Rev. D}, 80:054011, 2009.

\end{thebibliography}

\end{document}